\documentstyle[12pt,epsf]{article}
\newcommand{\beq}{\begin{equation}}
\newcommand{\eeq}{\end{equation}}
\newcommand{\beqa}{\begin{eqnarray}}
\newcommand{\eeqa}{\end{eqnarray}}

\begin{document}

\noindent Accepted for publication in Phys. Lett. {\bf B} \hfill KFA-IKP(TH)-1997-06

\hfill TK 97 07

\hfill hep-ph/9704377

\vspace{1in}

\begin{center}

{{\Large\bf Virtual photons in SU(2) chiral perturbation theory\\[0.2cm]
    and electromagnetic corrections to $\pi\pi$ scattering
    }}\footnote{Work supported
    in part by Deutsche Forschungsgemeinschaft (grant ME 864-11/1) and
    by funds provided by the Graduiertenkolleg "Die Erforschung 
    subnuklearer Strukturen der Materie".}

\end{center}

\vspace{.3in}

\begin{center}
{\large 
Ulf-G. Mei{\ss}ner$^\ddagger$\footnote{email: Ulf-G.Meissner@fz-juelich.de},
G. M\"uller$^\dagger$\footnote{email: mueller@itkp.uni-bonn.de},
S. Steininger$^\dagger$\footnote{email: sven@itkp.uni-bonn.de}}

\bigskip

$^\ddagger${\it Forschungszentrum J\"ulich, Institut f\"ur Kernphysik 
(Theorie)\\ D-52425 J\"ulich, Germany}

\bigskip

$^\dagger${\it Universit\"at Bonn, Institut f{\"u}r Theoretische Kernphysik\\
Nussallee 14-16, D-53115 Bonn, Germany}\\

\bigskip

\end{center}

\vspace{.6in}

\thispagestyle{empty} 

\begin{abstract}\noindent
We construct the generating functional of two flavor chiral perturbation
theory including the effects of virtual photons in the one loop approximation.
As an application, we calculate the electromagnetic corrections to the
elastic $\pi\pi$ scattering amplitude, in particular to its S--wave threshold 
parameters. Numerical estimates are given for the reaction $\pi^0
\pi^0 \to \pi^0 \pi^0$. These electromagnetic effects 
are found to be smaller than  the hadronic  two--loop
corrections for the scattering length $a_0$. The effective range $b_0$
increases by 36\% due to the unitary cusp.
\end{abstract}

\vfill

\pagebreak

\noindent {\bf 1.} 
Elastic pion--pion scattering is the purest reaction to test our understanding
of the spontaneous and the explicit chiral symmetry breaking in QCD. This 
reaction can be calculated to high precision within the framework of chiral
perturbation theory (CHPT) and is one tool to pin down the size of the scalar
quark condensate $B_0 = |\langle 0| \bar q q|0 \rangle |/ F_0$, with $ F_0$
the pion decay constant (in the chiral limit), and to give 
bounds on the up and down quark masses
(for a review, see~\cite{ulfrev}). CHPT is the effective field theory of the
standard model at low energies, which admits an expansion in small external
momenta and quark (pion) masses, with $q$ collectively denoting 
any one of these
small parameters. This so--called chiral expansion, which proceeds in steps
of  powers of $q^2$, can be mapped one--to--one on an expansion in the
number of pion loops and higher dimension local operators
\cite{wein79}. At present, elastic $\pi\pi$ 
scattering in the threshold region has been investigated at tree level ($q^2$)
\cite{wein66}, to one loop ($q^4$) \cite{gl84} and to two loop accuracy 
($q^6$) \cite{bcegs}. There exits also a dispersive analysis to order $q^6$
in the framework of generalized CHPT \cite{kmsf}.
Of particular interest are the S--wave
scattering lengths $a_0^I$, with $I = 0,2$ the total isospin of the two--pion
system, since they vanish in the chiral limit \cite{wein66}. Consider
as an example $a_0^0$ in CHPT: For the central values of the various
input parameters, the tree, one-- and two--loop results are $a_0^0
=0.156, \, 0.201$ and $0.217$, in order.\footnote{In particular,
  $F_\pi = 93.2\,$MeV and the charged pion mass are used. For a
  discussion on the parameter sensitivity, see \cite{bcegs}.}
While the one--loop corrections
are sizeable ($\sim 25\%$), the two--loop corrections are already considerably
smaller ($\sim 10\%$). Apart from these strong interaction
corrections, there are also electromagnetic contributions. In
particular, the charged to neutral pion mass difference,
which is an effect of the order of $(M_{\pi^+} -M_{\pi^0}) / M_{\pi^+} 
\sim \,$4\%, is almost entirely of
electromagnetic origin \cite{das}. It is therefore of importance to
get a handle on the electromagnetic corrections for the $\pi\pi$
scattering amplitude to further sharpen the theoretical
predictions. Virtual photons have already been included in three
flavor CHPT \cite{ru}\cite{nr} and various effects like e.g. the
violation of Dashen's theorem or the electromagnetic (em) corrections to
the decay $\eta \to 3 \pi$ \cite{bkw} have been calculated. Since 
$\pi \pi$ scattering can be described purely within SU(2) CHPT, we
construct here the generating functional including virtual photons to
one loop for the two flavor case. 
As proposed in \cite{ru}, we assign the chiral dimension one to the
electric charge, so the lowest order effective Lagrangian coupling the
virtual photons to the pions is of ${\cal O}(e^2) \sim {\cal O}(q^2)$. 
The difference to the three flavor case considered in 
refs.~\cite{ru}\cite{nr} comes largely from the fact that 
for two flavors there are less independent operators. This
reduced number of terms is the natural basis to consider purely pionic
processes. As an example, we then estimate the em corrections to
the $\pi \pi$ scattering amplitude, in particular to the threshold
parameters (scattering lengths, effective ranges). In this letter,
we concentrate on the reaction $\pi^0 \pi^0 \to \pi^0 \pi^0$ since
space forbids  a thorough discussion of the treatment of the infrared
divergences appearing in charged pion scattering. Previous work mostly
related to the leading em corrections involving at least one $\pi^+
\pi^-$  pair can be found in \cite{rs} \cite{wicky} \cite{mw}
\cite{rasche}. In most of these calculations, the pion mass difference
has not been included systematically.

\medskip

\noindent {\bf 2.} 
The effective field theory build of pions, collected in $U(x) = u^2(x)$,
photons $(A_\mu$) and other scalar
($s$), pseudoscalar ($p$), vector ($v_\mu$) and axial--vector
($a_\mu$) external sources starts at dimension two,
\beq \label{L2}
{\cal L}^{(2)} = - {1\over 4}F_{\mu\nu}F^{\mu\nu} - {\lambda \over 2} 
(\partial_\mu A^\mu )^2 + {F_0^2\over 4} \langle D_\mu U D^\mu
 U^\dagger + \chi U^\dagger + \chi^\dagger U \rangle + C \langle Q_R U
 Q_L U^\dagger \rangle \,\, , 
\eeq
where $\langle \,\, \rangle$ denotes the trace in flavor space, $F_{\mu
\nu} = \partial_\mu A_\nu - \partial_\nu A_\mu$ the photon field
strength tensor, $\lambda$ the gauge--fixing parameter (from here on,
 we work in the Lorentz gauge $\lambda = 1$), $D_\mu$ the generalized covariant
derivative,
\beq
D_\mu U = \partial_\mu U - i(v_\mu+a_\mu+QA_\mu)U
 +iU(v_\mu-a_\mu+QA_\mu) \,\, , 
\eeq
and  $Q$ the quark charge matrix,
\beqa
Q = {e \over 3} \left( \matrix { 2 & 0 \nonumber \\ 0 & -1
 \nonumber\\} \!\!\!\!\!\!\!\!\!\!\!\!\!\!\!\!\!\!  \right) 
= {e \over 6} \left( 3 \tau^3 + {\bf 1}  \right)
 \,\, .
\eeqa
In what follows, we work in the so--called $\sigma$--model gauge,
\beq
U(x) = \sigma(x) \, {\bf 1} + i \vec{\tau} \cdot \vec{\pi} (x) /F_0   
\,\, , \quad \sigma(x) = \sqrt{1 - \pi^2 (x)/F_0^2 } \,\, .
\eeq
The third term in Eq.(\ref{L2}) is the standard non--linear
 $\sigma$--model coupled to external sources (we neglect here
singlet axial components and set $\langle a_\mu \rangle  
=0$) \cite{gl84}. In particular, we have $\chi = 2B_0(s+ip)$ and the
external scalar source contains the quark mass matrix, $s(x) = 
{\rm diag}(m_u , m_d) + \ldots$. The last term in the dimension two
Lagrangian is the lowest order chiral invariant term one can
construct from pion and photon fields \cite{egpdr}. To make it
invariant under chiral SU(2)$_L\times$SU(2)$_R$ transformations, we have
introduced the spurions $Q_{L,R}$, which have the following
transformation properties,
\beq
Q_I \to g_I \, Q_I \, g_I^\dagger  \, , \quad 
g_I \in SU(2)_I \, , \quad I = L, R
\,\, . 
\eeq
At a later stage, one sets $Q_L = Q_R = Q$. The constant $C$ can be
calculated from the neutral to charged pion mass difference since this
term leads to $(\delta M^2)_{\rm em} = 2e^2C/F_0^2$. This 
identification is based upon the fact that the quark mass difference
$m_d - m_u$ only gives a tiny contribution to $M^2_{\pi^+} -
M^2_{\pi^0}$ \cite{glpr}. It can therefore be expected that these
quark mass effects are also tiny for the elastic $\pi \pi$ scattering 
amplitude. Consequently, we will work in the isospin
limit $m_d = m_u = \hat m$ in what follows.
For later convenience, we introduce the dimensionless constant
\beq \label{Z} 
Z = C / F_0^4 
= 0.89 \,\, , 
\eeq
with $F_0 = 88\,$MeV \cite{gl84}. As already stated before, we count
the external vector and axial--vector fields as well as the charge
matrices $Q, Q_L, Q_R$ as ${\cal O}(q)$ and the photon field as
${\cal O}(1)$. This has the advantage of a consistent power counting
between the strong and electromagnetic interactions, i.e. $e \sim q$
and one has terms of dimension two, four and so on. Here, dimension
two means either order $q^2$ or $e^2$ and similar at higher orders.
The construction
of the generating functional proceeds along standard lines making use
of heat kernel techniques for elliptic Euclidean differential
operators. We follow
here the approach outlined in \cite{ru} and combine the expansion
around the classical solutions for the meson and photons fields, $U =
U^{\rm cl} + i u \, \xi\, u/F_0 + \ldots$ and $A_\mu = A_\mu^{\rm cl} +
 \epsilon_\mu$, respectively,  
in one set of fluctuations variables $\eta = (\xi^1 , \ldots , \xi^3,
\epsilon^0, \ldots ,\epsilon^3)$. One then expands the generating
functional 
\beq
\exp\bigl[i {\cal Z}(s,p,v_\mu,a_\mu) \bigr] = {\cal N} \int [dA_\mu]
[dU] \exp\biggl\{ i \int d^4x \, ( {\cal L}_2 + {\cal L}_4 ) \biggr\}
\eeq
up to second order in the fluctuations $\eta$.
Using dimensional regularization, the
divergences can be extracted in a straightforward manner. For the two
flavor case at hand, the em part of the dimension four counterterm
Lagrangian takes the form
\beq \label{L4}
{\cal L}^{(4)}_{\rm em} = \sum_{i=1}^{13} k_i \, {\cal O}_i \,\, , 
\eeq
with the ${\cal O}_i$ monomials in the fields of dimension four. The
low--energy constants $k_i$ absorb the divergences in the standard
manner,
\beqa \label{L4em}
k_i &=& \kappa_i \, L + k_i^r (\mu ) \, \, , \\
L &=& {\mu^{d-4}\over 16\pi^2} \biggl\{ {1 \over d-4} - {1\over 2}
 \biggl[ \ln(4\pi ) + \Gamma '(1) +1 \biggr] \biggr\} \,\, ,
\eeqa
with $\mu$ the scale of dimensional regularization and $d$ the number
of space--time dimensions. 
The explicit expressions for the operators
${\cal O}_i$ and their $\beta$--functions $\kappa_i$ are collected in 
table~1. The $k_i^r (\mu)$ are the renormalized, finite and
scale--dependent low--energy constants. These can  be fixed by data or
have to be estimated with the help of some model. 
\renewcommand{\arraystretch}{1.3}

\begin{table}[t]
$$
\begin{tabular}{|r|c|c|} \hline
i  &     ${\cal O}_i$  &          $\kappa_i$ \\ \hline
1 & $ \langle QUQU^\dagger \rangle^2   $
& $ 3/2 + 3 Z + 12 Z^2  $ \\
2 & $ \langle QUQU^\dagger \rangle \, 
                   \langle D_\mu U  D^\mu U^\dagger \rangle $
& $ 2 Z  $ \\
3 & $ \langle U^\dagger D_\mu U Q \rangle^2 + \langle D_\mu U  U^\dagger
Q \rangle^2 $ & $ -3/4 $ \\
4 & $ \langle U^\dagger D_\mu U Q \rangle \, \langle D^\mu U  U^\dagger
Q \rangle $ & $ -2 Z $ \\
5 & $ \langle [Q, [D_\mu, Q]]  U^\dagger D^\mu U\rangle -  
\langle [Q, [D_\mu , Q]] D^\mu U U^\dagger \rangle
$ & $ -1/4 $ \\
6 & $ \langle [D_\mu , Q] U [D^\mu , Q] U^\dagger \rangle $ & $0$ \\
7 & $ \langle QUQU^\dagger \rangle \,  \langle \chi U^\dagger 
                                    + U \chi^\dagger\rangle   $
& $ 1/4 + 2 Z   $ \\
8 & $ \langle (U^\dagger \chi - \chi^\dagger U ) \,  
( U^\dagger Q U Q - Q U^\dagger Q U ) \rangle
$ & $ 1/8 - Z  $ \\
9 & $\langle Q^2 \rangle \,\langle QUQU^\dagger \rangle   $
& $ -3 - 3 Z/5 - 12 Z^2 /5 $ \\
10 & $ \langle Q^2 \rangle \,  \langle D_\mu U  D^\mu U^\dagger \rangle $
& $ -27/20 - Z / 5   $ \\
11 & $ \langle Q^2 \rangle \,  \langle \chi U^\dagger 
                                    + U \chi^\dagger\rangle   $
& $ -1/4 - Z / 5   $ \\
12 & $ \langle [D_\mu , Q_R][D^\mu , Q_R] +  [D_\mu , Q_L][D^\mu , Q_L]
\rangle  $ & $0$ \\ 
13 & $\langle Q^2 \rangle^2 $
& $3/2 - 12 Z / 5 + 84 Z^2 / 25 $ \\
\hline
\end{tabular}
$$
\caption{Counterterms and their $\beta$--functions.}
\end{table}
\noindent
As it is the case for the hadronic LECs in the
two--flavor case \cite{gl84}, one can introduce scale--independent 
couplings, called $\bar{k}_i$, via
\beq
k_i^r (\mu ) = {\kappa_i \over 32 \pi^2}\,\,  \biggl[\,\bar{k}_i + \ln 
\frac{M_{\pi^0}^2}{\mu^2}\,\biggr] \quad .
\eeq
Notice that we choose the neutral pion mass as the reference
scale. This is natural since the neutral pion mass is almost entirely 
a hadronic effect, in contrast to the charged one.
Some remarks concerning these results are in order. To arrive
at the structures given in the table, we have made use of the equations of
motions for the classical $U$ field as well as of the Cayley--Hamilton
matrix relations. This minimizes the number of independent terms. 
The last term in the table $\sim \langle Q^2 \rangle^2$ does not
contain any pion field and is only needed for the complete
renormalization, i.e. it does not influence physical processes.
The same applies to  ${\cal O}_{12}$.
Similarly, the operators ${\cal O}_{9,10,11}$ only lead to
renormalizations of $C$, $F_0$ and $M_0$, with $M_0^2 = 2 \hat m B_0$
the leading term in the quark mass expansion of the pion mass.
A typical difference between SU(2) and SU(3) is the operator ${\cal
O}_{11}$, which has a vanishing $\beta$--function in SU(3)
\cite{ru}. Note furthermore that symmetry does allow for terms
which are odd in powers of $Q$, like e.g. a term $\sim \langle Q^2 U Q
U^\dagger \rangle$. Such terms are, however, unphysical since the 
electric charge always appears in even powers for physical processes.
Consequently, we did not further study such terms. 
We also have repeated the SU(3) calculation and found that
the $\beta$--functions for the terms $K_{15,16,17}$ are incorrectly
given in the existing literature. The correct ones read (in the
notation of Urech \cite{ru})\footnote{
Res Urech has kindly informed us that he has checked his coefficients
$\Sigma_{15,16,17}$ finding agreement with our results.}
\beq
\Sigma_{15} = 3/2 + 3Z + 14Z^2 \, , \quad
\Sigma_{16} = -3 - 3Z/2 - Z^2 \, , \quad
\Sigma_{17} = 3/2 - 3Z/2 + 5Z^2 \, \, .
\eeq
This difference in the terms $\sim Z^2$
can be traced back to the fact that the operator $\sigma^{ab}$,
which appears in the four--dimensional Euclidean one--loop functional,
has to be symmetrized. This was not done in the part $\sim C$ in
\cite{ru}, i.e. when going from Eq.(15) to Eq.(21) in that paper.

\medskip

\noindent {\bf 3.} Before evaluating the pertinent em corrections, 
we collect here some basic definitions
concerning the elastic $\pi\pi$ scattering amplitude (in the threshold 
region). Consider the process $\pi^a (p_a) + \pi^b (p_b) \to \pi^c (p_c)
+ \pi^d (p_d)$, for pions with isospin $'a,b,c,d\,'$ and momenta $p_{a,b,c,d}$.
The corresponding Mandelstam variables are $s = (p_a + p_b)^2$, 
$t = (p_a - p_c)^2$, $u = (p_a - p_d)^2$ with $s+t+u = M^2_{\pi^a} + 
M^2_{\pi^b} + M^2_{\pi^c} + M^2_{\pi^d}$. The scattering amplitude can 
be expressed in terms of a single function, denoted
$A(s,t,u)$,\footnote{This is  correct in the isospin--conserving
  case. The more general form of the amplitude in case of isospin
  violation will be given in \cite{su}.}
\beq
T^{cd;ab} = A(s,t,u) \, \delta^{ab}\delta^{cd} + 
 A(t,s,u) \, \delta^{ac}\delta^{bd} +  A(u,t,s) \, \delta^{ad}\delta^{bc} 
\,\, .
\eeq
The chiral expansion of $A(s,t,u)$ takes the form
\beq
A(s,t,u) = A^{(2)}(s,t,u) + A^{(4)}(s,t,u) + {\cal O}(q^6) \,\,\, ,
\eeq
where $ A^{(m)}$ is of order $q^m$ and the symbol ${\cal O}(q^6)$ denotes
terms like $s^3, s^2 t, s^2 e^2, s e^4, \ldots\,$. The form of 
$A^{(2)}(s,t,u)$ was first given by Weinberg \cite{wein66} and the 
next--to--leading order terms by Gasser and Leutwyler  \cite{gl84}.
For convenience, one splits this contribution as $A^{(4)}(s,t,u) =
B(s,t,u) + C(s,t,u)$, where $B(s,t,u)$ collects the unitarity corrections
and $C(s,t,u)$ the (real) tree and tadpole contributions.
To evaluate the one--loop graphs, we introduce the modified pion propagator 
\beq \label{prop}
\Delta_\pi^{ab} (\ell)
= {i \delta^{ab}\over [\ell^2 - M_{\pi^0}^2 
- \delta M^2  \, ( 1- \delta^{3a} \, )]} \,\, ,
\quad \delta M^2 = {2 e^2 C \over F_0^2} \,\, ,
\eeq
with $\ell$ the pion four--momentum  and $'a,b\,'$ isospin indices.
The form of Eq.(\ref{prop}) is due to the fact 
that in this gauge, the operator $\sim C\langle
QUQU^\dagger\rangle$ only contributes to terms which are 
quadratic in pion fields. To evaluate the unitarity corrections, i.e.
the diagrams which give rise to the imaginary part of the scattering
amplitude, we introduce a generalization of the commonly used ``bubble
function'' $\bar J$,
\beq
\bar J (q^2,M^2,A) \equiv {1\over 16\pi^2} \biggl[ {\sigma\over 2} \biggl(
\ln \frac{\sigma_- - 1}{\sigma_- + 1} + \ln 
\frac{\sigma_+ - 1}{\sigma_+ + 1}  
\biggr) + 2 \biggr] \,\,   ,
\eeq
with 
\beqa
\sigma &=& \sqrt{1 - \frac{4M^2}{s} + \frac{A}{s} \biggl( \frac{A}{s}+2
\biggr)} \,\, , \\
\sigma_- &=&  \sqrt{1 - \frac{4M^2}{s} \biggl(  \frac{A}{s} + 1 
\biggr)^{-2}}\,\,  , \quad
\sigma_+ = \sqrt{1 - \frac{4M^2}{s} \biggl(1 - \frac{A}{M^2}\biggr)
\biggl(  \frac{A}{s} - 1 \biggr)^{-2}} \,\, ,
\eeqa
with $A$ a  quantity of dimension [mass$^2$] like e.g.  
$\delta M^2$. For $A = 0$, one
recovers  the standard form of $\bar J$ \cite{gl84} since then 
$\sigma = \sigma_- = \sigma_+ = (1 - 4M^2 / s)^{1/2}$. 
All loop integral can be expressed in 
terms of $\bar J (q^2,M^2,A)$ and some polynoms (modulo logarithms). 
In terms of physical processes, we have five reaction channels
\beqa \label{chan}
&&{\rm (a)} \quad \pi^0 \, \pi^0 \to  \pi^0 \, \pi^0 \,\, , \qquad
{\rm (b)} \quad \pi^+ \, \pi^- \to  \pi^0 \, \pi^0 \,\, , \qquad
{\rm (c)} \quad \pi^+ \, \pi^- \to  \pi^+ \, \pi^- \,\, , \nonumber \\
&&{\rm (d)} \quad \pi^0 \, \pi^+ \to  \pi^0 \, \pi^+ \,\, , \quad\,\,\,
{\rm (e)} \quad \pi^+ \, \pi^+ \to  \pi^+ \, \pi^+ \,\, . 
\eeqa
For comparison with the data, one decomposes $T^{cd;ab}$ into amplitudes
of definite total isospin $(I=0,1,2)$ and projects out partial--wave
amplitudes $T_l^I (s)$,
\beq
T_l^I (s) = {\sqrt{1-4M_\pi^2 / s}\over 2i} \biggl[ \exp \bigl\{
2 i [ \delta_l^I (s) + i \eta_l^I (s) ]\bigr\} - 1 \biggr] \,\, ,
\eeq
with $ s = 4(M_\pi^2 + q^2)$ and $q$ the pion momentum
in the c.m. system.\footnote{This holds only 
for the equal mass case. The generalization to the unequal mass 
case is obvious.} Furthermore,  $l$ denotes the total angular momentum of the 
two--pion system. The phase shifts $\delta_l^I (s)$ are real and the
inelasticities $\eta_l^I (s)$ set in at four--pion threshold. Below
$\bar K K$ threshold, $s \simeq 1\,$GeV$^2$, 
they are negligible and will be ignored in what
follows. Near threshold, the partial--wave amplitudes take the form
\beq
{\rm Re} \, T_l^I (s) = q^{2l}\bigl\{ a _l^I + q^2 \, 
b_l^I + {\cal O}(q^4) \bigr\} \,\, .
\eeq
The coefficients $a _l^I$ are called scattering lengths, the $b_l^I$
are the range parameters. In the following, we will concentrate on these
quantities. 

\medskip

\noindent {\bf 4.} We are now in the position to evaluate the em corrections
to the elastic $\pi \pi$ scattering amplitude. To one--loop order, one
has of course the standard strong interaction graphs, i.e. tree
graphs at orders $q^2$ and $q^4$ as well as one--loop digrams at
${\cal O}(q^4)$. In the $\sigma$--model gauge, these can be calculated
according to standard methods, the only difference being the modified
pion propagator, Eq.(\ref{prop}). The corresponding Feynman diagrams 
are shown in fig.~1a,b,c,d. There are two tree graphs at order $e^2$. 
Diagram a) contributes to the charged pion mass shift and b) vanishes
for the reasons discussed above. That is also the reason why there is
no one--loop graph like c) with the insertion on the four--pion
vertex. The next seven graphs in fig.1 are of order ${\cal O}(e^2 q^2)$.
From the irreducible photon loop diagrams (fig.1e,f,g) only
the first one is non--vanishing  since one has no $\gamma 4\pi$ 
and $2\gamma 4\pi$ vertices in this particular gauge (this point was
also stressed by Gasser$\,$\cite{fras} and Ecker$\,$\cite{erice}).
Clearly, graph h) are only gives rise to wave function
renormalization and i) vanishes in dimensional regualarization. 
Finally, the counterterms with exactly one insertion from
${\cal L}^{(4)}_{\rm em}$ are depicted in figs.1j,k. From these two,
only graph$\,$1j gives rise to a genuine em correction to the $\pi\pi$
scattering amplitude. For charged pions, there are in addition photon
exchange graphs as shown in fig.2. At tree level ${\cal O}(e^2)$ one has
the one--photon exchange diagram. To one--loop order $e^4$ and $e^2 q^2$, there
are six topologically different  graphs. While diagram~2a is of order
$e^2 q^2$, the others are of ${\cal O}(e^4)$. In particular, there
are three types of two--photon exchange graphs with one and two
intermediate pion propagators, figs.2b,c,d, in order. The
last two diagrams in fig.2 are simply vertex and self--energy
corrections. As one easily convinces oneself, the photon exchange
graphs like 1e, 2a, $\ldots$, 2f lead to IR divergences. These are
cancelled in the standard fashion by considering the radiation of very
soft photons in the initial and final states.\footnote{A lucid
discussion of such infrared effects is given in chapter 13 
of ref.\cite{weinI}. A detailed study of radiative four--meson
amplitudes in CHPT can be found in \cite{daeih}. }
For comparison with experiment, one can also use the standard Gamov
factors \cite{gamov} to remove the Coulomb enhancement in the initial
and the final state for charged pions. What we are really after are
the electromagnetic effects once these "kinematical" em effects are removed. 
Space forbids here to discuss these matters in detail and we refer
to ref.\cite{su} for a comprehensive treatment. 
In what follows, we consider only the scattering of neutral pions, 
i.e. the reaction (a) in Eq.(\ref{chan}).

\medskip

\noindent {\bf 5.} For $\pi^0\pi^0 \to \pi^0 \pi^0$, we have $s+t+u =
4 M_{\pi^0}^2$ and the leading order Weinberg term reads
\beq \label{A2}
A^{(2)} (s,t,u) = {s - M_{\pi^0}^2 \over F_\pi^2}
\eeq
in terms of the physical values $M_{\pi^0} = 134.97\,$MeV and $F_\pi =
92.5\,$MeV.\footnote{Note that the extraction of this value for
$F_\pi$ includes one--loop radiative corrections \cite{barry}.} The
respective shifts from the lowest order values $M$ and $F_0$ are
accounted for in the next--to--leading order contribution $C(s,t,u)$.
It is important to stress that the lowest order amplitude is therefore
not affected directly by the em corrections, only indirectly through
the pion mass shift. This result is, of course, well--known \cite{fras}. 
The unitarity corrections take a form similar to the purely strong
interaction  results of \cite{gl84},   
\beqa \label{Bstu}
B(s,t,u) = {1 \over 6 F_\pi^4} 
           && \!\!\!\!\!\!\!\!\!\!\! \biggl[6(s-M_{\pi^0}^2)^2
                  \bar{J}(s,M_{\pi^+},0)
                  -3(s^2-4sM_{\pi^0}^2+3M_{\pi^0}^4)
                  \bar{J}(s,M_{\pi^0},0) \nonumber \\
           && + \, [t(t-u)-2M_{\pi^0}^2t+4M_{\pi^0}^2u-2M_{\pi^0}^4]
                  \bar{J}(t,M_{\pi^0},0)
           \nonumber \\
           && + \,  [u(u-t)-2M_{\pi^0}^2u+4M_{\pi^0}^2t-2M_{\pi^0}^4]
                  \bar{J}(u,M_{\pi^0},0)\biggr]
\eeqa
The next--to-leading order tree and tadpole contributions read
\beqa \label{Cstu}
&& \!\!\!\!\! \!\!\!\!\! \!\!\!\!\!
  C(s,t,u) = 
 {1 \over 96\pi^2 F_\pi^4} 
     \biggl[ 2 (\bar{l}_1-\frac{4}{3})(s-2M_{\pi^0}^2)^2   
             + (\bar{l}_2-\frac{5}{6})\{s^2+(t-u)^2\} 
  - 3M_{\pi^0}^4 \, \bar{l}_3 \nonumber \\
&&  \qquad\qquad\qquad\qquad      + 12 M_{\pi^0}^2 (s - M_{\pi^0}^2) \,
                   \bar{l}_4 - 12M_{\pi^0}^2 s+  15M_{\pi^0}^4              
              \biggr] \nonumber \\
&&  \qquad -\frac{1}{16\pi^2F_\pi^4}\,\, \biggl\{ \, 
    \ln\frac{M_{\pi^+}^2}{M_{\pi^0}^2}
    \biggl[ M_{\pi^+}^2 (3s-4 M_{\pi^0}^2) + (s-M_{\pi^0}^2)^2
    \biggr]            \nonumber \\  
&& \qquad\qquad\qquad\qquad    
    + e^2 \, F_\pi^2 \, (3 s - 4M_{\pi^0}^2) \, \biggl[ 
   {3 \over 2}\bar{k}_3 + 2Z \bar{k}_4 \, \biggr] \, \biggr\}  \,\, . 
\eeqa
The last two terms deserve some discussion. First, the term 
$\sim \ln(M_{\pi^+}^2/M_{\pi^0}^2)$ is due to the fact that we normalize
the scale--independent SU(2) low--energy constants $\bar{\ell}_i$ 
to the neutral pion mass and that in the loops one has neutral as well
as charged pion pairs propagating. 
The last term in Eq.(\ref{Cstu}) is the novel em contribution 
from ${\cal L}^{(4)}_{\rm em}$. To arrive at these results, we
have used the low--energy expansions of 
the pion mass and the pion decay constant,\footnote{In this letter,
we identify the neutral with the charged pion decay constant. A more thorough
discussion of the em effects on the AA correlators is given in \cite{su}.}
\beqa
M_\pi^2 &=& M_0^2 \, \Bigg\{ 1 -\frac{M_0^2}{32\pi^2 F_0^2} \, \bar{l}_3
 + \frac{M_+^2}{16\pi^2F_0^2} \ln\frac{M_+^2}{M_0^2}
+\frac{5e^2}{72\pi^2}
\biggl[\biggl(2Z+{1\over 4}\biggr) \, \bar{k}_7 -
\biggl( {Z\over 5}+ {1\over 4}\biggr) \, \bar{k}_{11}\biggr]
\nonumber \\
&& \qquad\qquad - \frac{5e^2}{72\pi^2}\biggl[2Z \, \bar{k}_2-
\biggl({Z\over 5}+ {27 \over 20} \biggr) \, \bar{k}_{10}  \biggr]
-\frac{e^2}{32\pi^2}\biggl[3 \, \bar{k}_3 + 4Z \,\bar{k}_4\biggr]
\Bigg\}  \,\, ,  \\
F_\pi &=& F_0 \, \Bigg\{ 1 + \frac{M_0^2}{16\pi^2F_0^2} \, \bar{l}_4 
-  \frac{M_+^2}{32\pi^2F_0^2} \ln\frac{M_+^2}{M_0^2} \nonumber \\
&& \qquad\qquad    +  \frac{5e^2}{144\pi^2}
\biggl[2Z \, \bar{k}_2- \biggl( {Z\over 5}+ {27\over 20}\biggr)
\, \bar{k}_{10}\biggr]
+ \frac{e^2}{64\pi^2}\biggr[3\, \bar{k}_3+4Z\, \bar{k}_4\biggr]
\Bigg\} \,\, , 
\eeqa
where $M_+^2 = M_0^2 + \delta M^2$, see Eq.(\ref{prop}).
Notice that in $B(s,t,u)$ and $C(s,t,u)$ we have set $F_0 = F_\pi$,
$M_0 = M_{\pi^0}$ and $M_+ = M_{\pi^+}$
since these differences are of order $q^2$ and thus beyond the accuracy
of the calculation presented here. As a check, we recover the result
for $A^{(2)} +A^{(4)}$ of \cite{gm} for one common pion mass and setting
$e = 0$. From the amplitude given in
Eqs.(\ref{A2},\ref{Bstu},\ref{Cstu}), it is straightforward to
calculate the pertinent scattering lengths $a_0 (00;00)$ and effective
ranges $b_0 (00;00)$ numerically or analytically \cite{su}. These are
related to the ones in the isospin basis via
\beq
a_0 (00;00) = {1\over 3}\, a_0^0 + {2\over 3}\, a_0^2 \,\, , \quad
b_0 (00;00) = {1\over 3}\, b_0^0 + {2\over 3}\, b_0^2 \,\, .  
\eeq
One can show analytically that for the process $\pi^0 \pi^0 \to \pi^0
    \pi^0$, the operators $\sim \bar{k}_i$ do not contribute.

For the numerical evaluation, we use the central values of the strong
low--energy constants $\bar{\ell}_i$ from \cite{bcg}, $\bar{\ell}_1 = -1.7$,
$\bar{\ell}_2 = 6.0$ together with $\bar{\ell}_3 = 2.9$, $\bar{\ell}_4 = 4.3$
and the values for $F_\pi$, $M_{\pi^0}$ and
$M_{\pi^+}$ given above. The corresponding S--wave threshold parameters are
given in table~2 in comparison to the strong one--loop results and the
experimental data. 

\smallskip
 
\renewcommand{\arraystretch}{1.4}
\begin{table}[bht] 
\begin{center}

\begin{tabular}{|c|c|c|c|c|c|c|}
    \hline
    & $a_0 (00;00) $ & $b_0 (00;00) $  \\ 
    \hline
  $e = 0$     &  0.0360   & 0.0302  \\    
  $e \neq 0$  &  0.0340   & 0.0412  \\    
    \hline    
  Exp. & $0.056\pm 0.027$ \protect{\cite{ochs}}\protect{\cite{bkmppn}} 
       & $0.029\pm 0.044$ \protect{\cite{nagels}} \\ 
         \hline
  \end{tabular}
\end{center}
\caption{S--wave threshold parameters with em corrections at one loop 
($e \neq 0$) compared to the hadronic one--loop results ($e = 0)$ 
in units of the inverse neutral pion mass. The data should only be
considered indicative since they mostly stem from processes involving
charged pions. The uncertainties are added in quadrature. } 
\end{table}

\noindent For the scattering length $a_0$ the effect of the em corrections
is of the order of  5\%, still approximately a factor of two smaller than the
strong two--loop correction \cite{bcegs}.  For the range parameter
$b_0$, however, we observe an 36\% increase. This is due to
unitarity cusp at $s_0 = 4 M_{\pi^+}^2$, which is expected to scale 
as $\sqrt{ M_{\pi^+}^2 -  M_{\pi^0}^2} /M_{\pi^+} \simeq 26$\%. 
The dominant isospin--violating effect is thus entirely given through the
charged to neutral pion mass difference, similar to the case of the
electric dipole amplitude in neutral pion photoproduction off nucleons
\cite{bkmzpc}.

\medskip

\noindent {\bf 6.} To summarize, we have constructed the generating
functional for two--flavor chiral perturbation theory including the
effects of virtual photons in the one--loop approximation. Counting
the electric charge as a small momentum, there are in total 13 terms
contributing to the em Lagrangian at next--to--leading order (some
of these are only needed for renormalization). As an application, we have
considered the em corrections to the elastic $\pi\pi$ scattering
amplitude and given numerical estimates for $\pi^0 \pi^0 \to \pi^0
\pi^0$. The charged to neutral pion mass difference
produces an pronounced effect on the S--wave effective range.
In a forthcoming publication, we will present results also
for the channels involving charged pions \cite{su} and include the
effects due to the quark mass difference $\sim m_d - m_u$. 

\vskip 1cm

\section*{Acknowledgements}

We are grateful to J\"urg Gasser and Kolya Nikolaev for very helpful 
comments and \\ V\'eronique Bernard for some independent checks. We
thank Marc Knecht for pointing out an inconsistency in the original version.


\newpage

\section*{Figures}

\vskip 1.5cm

\begin{figure}[h]
   \vspace{0.5cm}
   \epsfysize=11.9cm
   \centerline{\epsffile{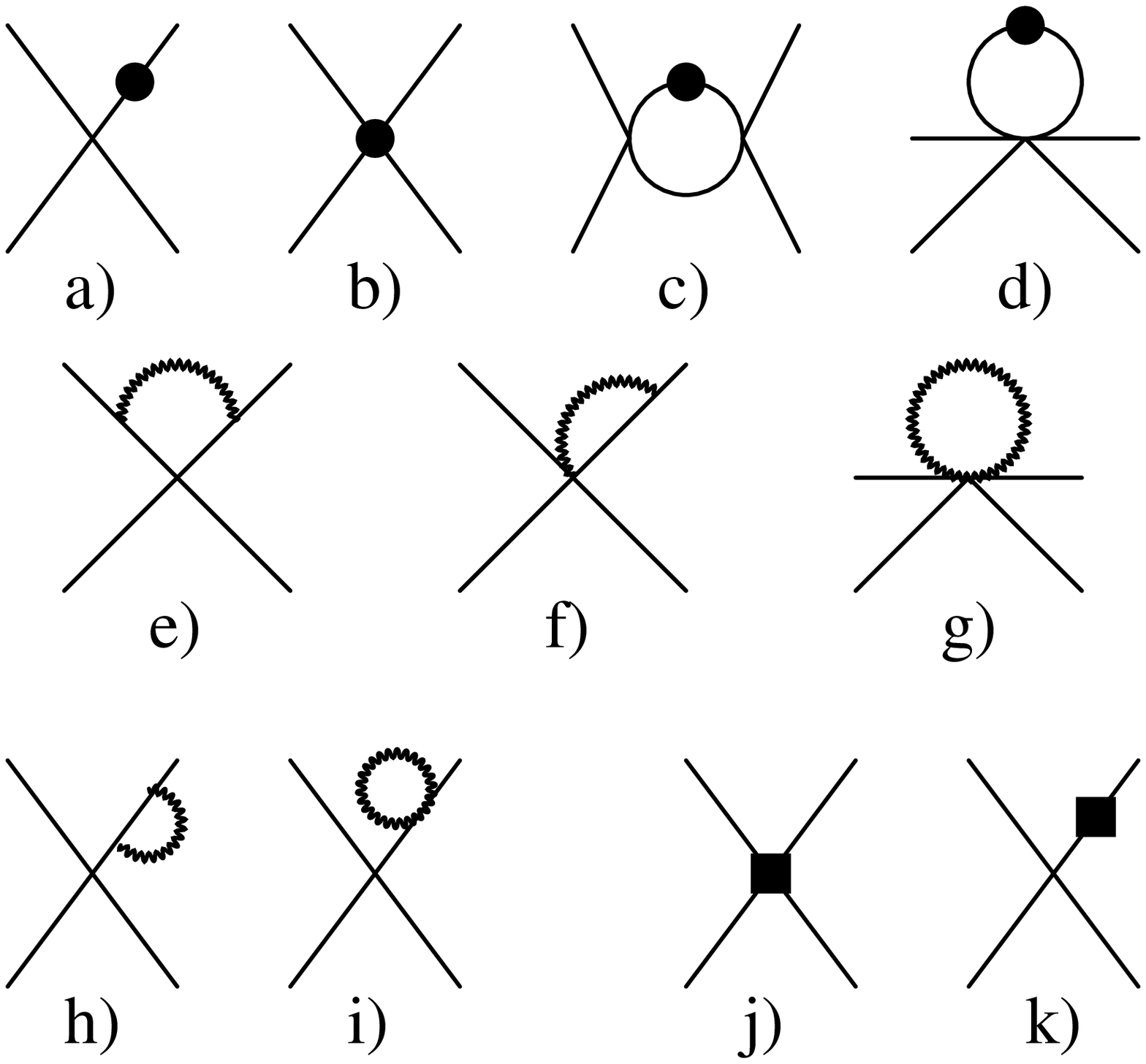}}
   \vspace{2cm}
   \centerline{\parbox{15cm}{\caption{\label{fig1}
   Graphs contributing to the em corrections to elastic $\pi\pi$
   scattering. Solid and wiggly lines denote pions and photons, in
   order. The heavy dot and the box refer to insertions from the em
   counterterms at order $e^2$ and $e^4$, respectively. Graphs b),
   f) and g) vanish in the $\sigma$--model gauge. i) vanishes in 
   dimensional regularization. Crossed graphs are not shown.
  }}}
\end{figure}


$\,$

\vskip 3cm

\begin{figure}[h]
   \vspace{0.9cm}
   \epsfysize=15cm
   \centerline{\epsffile{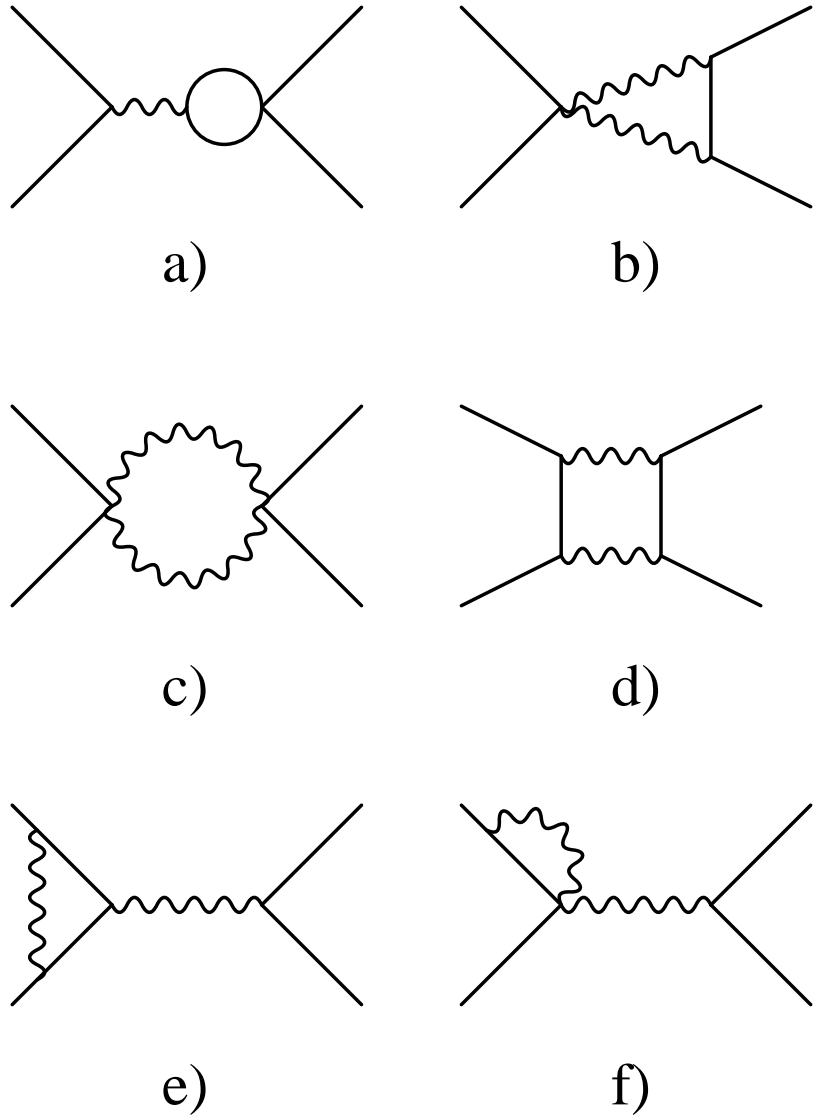}}
   \vspace{2cm} 
   \centerline{\parbox{15cm}{\caption{\label{fig2}
   One and two--photon exchange graphs contributing to the em 
   corrections for elastic scattering of charged pions at one loop. 
   While diagram a) is of order $e^2q^2$, the others are of ${\cal
   O}(e^4)$. Crossed graphs are not shown.  For notations, see fig.1.
  }}}
\end{figure}

\end{document}